# ANALYSES AND PERFORMANCE OF TECHNIQUES PAPR REDUCTION FOR STBC MIMO-OFDM SYSTEM IN (4G) WIRELESS COMMUNICATION

Leila Sahraoui, Djmail Messadeg, Nouredinne Doghmane

Department of Electronics Faculty sciences of Engineering
University Baji Mokhtar, Annaba bp 12 el hadjar, Algeria

*ABSTRACT*

*An OFDM system is combined with multiple-input multiple-output (MIMO) in order to increase the diversity gain and system capacity over the time variant frequency-selective channels. However, a major drawback of MIMO-OFDM system is that the transmitted signals on different antennas might exhibit high peak-to-average power ratio (PAPR).In this paper, we present a PAPR analysis reduction of space-time-block-coded (STBC) MIMO-OFDM system for 4G wireless networks. Several techniques have been used to reduce the PAPR of the (STBC) MIMOOFDM system: clipping and filtering, partial transmit sequence (PTS) and selected mapping (SLM). Simulation results show that clipping and filtering provides a better PAPR reduction than the others methods and only SLM technique conserve the PAPR reduction in reception part of signal.*

## 1. INTRODUCTION:

Orthogonal Frequency Division Multiplexing (OFDM) is multi-carrier modulation (MCM) scheme technique used for 4th Generation (4G) wireless communication. This technique with high-speed data transmission is used in mobile communication, Digital terrestrial mobile communication, Digital Audio Broadcasting (DAB) and Digital Video Broadcasting terrestrial (DVB-T). OFDM has many advantages such as High spectral efficiency, immunity to inter-symbol interference and capability of handling very strong multipath fading [1]. Despite its competitive position attributes, OFDM signals are characterized by very high levels of Peak to average power (PAPR Ratio). This peculiarity conduit OFDM signals to be very sensitive to non-linearities of the analog components of the transceiver, in particular those of the high power amplifier (HPA) at transmission. A HPA is designed to operate in its saturation region which corresponds to a high yield of the region.
However, in this area, the HPA nonlinear behavior severe. These nonlinearities are sources of in-band (IB) distortions that can both reduce the performance in terms of Bit Error Rate link (BER) [1]. For all these reasons, reducing the PAPR of OFDM signals is increasingly being considered to be very important to preserve the cost-effectiveness advantages of OFDM in practical systems. MIMO system is the use of multiple antennas at both the transmitter and receiver to enhance communication performance. OFDM combined with MIMO technology with the block of space-time coding (STBC) represents an interesting candidate for mobile communication systems due to its ability to withstand high speeds, high capacity and robustness to multipath fading [2]. However, as with single-input single-output OFDM (SISOOFDM), one of the major drawbacks of STBC MIMO-OFDM is the high peak-to-average power ratio (PAPR) of the signals transmitted on different antennas. A straightforward way to reduce the PAPR in STBC MIMOOFDM systems would be to apply the PAPR reduction approach proposed for SISO-OFDM systems [3] to each antenna separately and the side information (SI) required to recover the signal successfully at the receiver would increase proportionally with the number of antennas [4], [5]. A space-time coding system (STBC) with MIMO-OFDM is deployed for transmit diversity and secure means for data propagation in the scenario where the mobility is required for the data transmission. Is less a high PAPR always accompanies MIMO-OFDM systems. To remedy this problem, there are now a multitude of techniques for reducing PAPR have been proven in many applications of digital communications such as wireless networking son (Wi-Fi), the (WiMAX), HiperLAN and recently in the DVBT2. Many methods have been proposed to mitigate the OFDM PAPR by acting on the signal itself [2], [6]. The simplest ones use clipping and filtering techniques [4], [5]. However, these methods may increase the BER of the system since clipping is a nonlinear process [7].

Alternative methods are based on coding [8] and others on Multiple Signal Representation (MSR) techniques: Partial Transmit Sequence (PTS) [9], Selective Mapping (SLM) [10]. The main drawback of these methods is that a Side Information (SI) has to be transmitted from the transmitter to the receiver to recover the original data, which results in some loss of throughput efficiency [7].

Among these techniques, we propose to use a combination of STBC coding techniques (Alamonti) in MIMO-OFDM, with reduction techniques to better discriminate the PAPR reduction induced by the most widely used methods are clipping and filtering, selective mapping (SLM) and partial transmit sequences (PTS).

Our study is based on the PAPR reduction in transmission. Two approaches were made: Reduce the PAPR issue and compare the shape of the complementary cumulative distribution function (CCDF) curve of the PAPR for transmission and reception for each of the three methods mentioned above.

## 2. STBC MIMO-OFDM system:

We can write the complex vector of size $N$

$$X_\kappa = X_0, X_1, .......... X_{N-1}$$

Where, $X_k$ is the complex value carried by the k[th] subcarrier. The OFDM symbol can be written as:

$$x(t) = \frac{1}{\sqrt{N}} \sum_{k=0}^{N-1} X_k e^{j2\pi K f_0 t}, 0 \leq t \leq T \quad (1)$$

Where T is the symbol interval, and $f_0 = 1/T$ is the frequency spacing between adjacent subcarriers.

Replacing t=n $T_b$, where $T_b$ =T/N, gives the discrete time version denoted by

$$x(n) = \frac{1}{\sqrt{N}} \sum_{k=0}^{N-1} X_k e^{j2\pi kn/LN}, n = 0,1,......NL-1 \quad (2)$$

Where, L is the oversampling factor.

Fig. 1 illustrates the general block diagram of a STBC MIMO-OFDM system. Baseband modulated symbols are passed through serial-to-parallel (S/P) converter which generates complex vector of size *N*. Then the complex vector, *X* is then passed through the STBC encoder. Both these sequences are then passed through each IFFT block for antenna 1 and antenna 2 respectively.

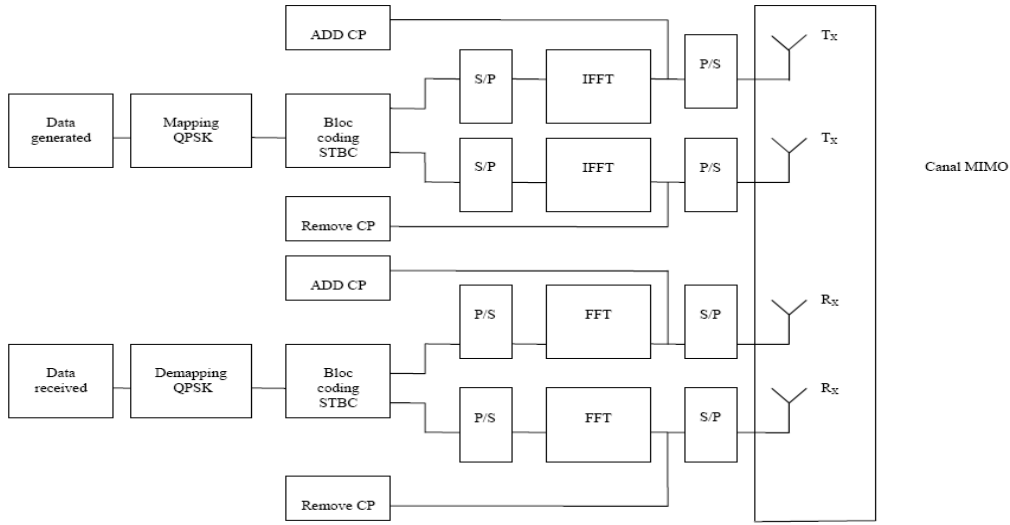

Figure 1. Structure of STBC MIMO-OFDM

## 2.1. Alamouti Space-Time Block Code (STBC):

Increasing the transmission rate and providing robustness to channel conditions are nowadays two of the main research topics for wireless communications. Therefore, a significant interest of late has been to develop systems that offer both high capacity and high data speed. Indeed, much effort is done in the area of Multiple Input Multiple Output (MIMO) systems by using several antennas either at the transmitting side or at the receiving side. We can exploit space and time diversity by using Space time codes such as the famous Alamouti code In multi antennas communication systems, investigations of *Nt* transmit and *N r* receive antennas showed that the capacity of such systems increases linearly with the minimum of *Nt* and *Nr*. High data rates are obtained by simultaneously sending signals from several transmit antennas [11].

The space-time coding technique is essentially a two dimensional space and time processing method. While multiple antennas both for transmission and reception are used to improve wireless communication systems capacity and data rate in space-domain, in time-domain, different signals can be transmitted at different time slots using the same antenna at the same time. Correlation of time and space is introduced between signals which are transmitted by different antennas so that the receiver antennas can realize diversity reception.

Therefore, space-time coding is especially meant for higher coding gain without using more bandwidth which effectively enhances capacity of wireless systems [12], [13], [14].

In our paper, we have used the Alamouti space time coding to achieve spatial diversity in MIMO systems. In Alamouti space time block coding (STBC), every two continuous transmit symbols, $s_1$ and $s_2$, are coded into a transmit symbol matrix as follows:

$$s = \begin{bmatrix} s_1 & -s_2^* \\ s_2 & s_1^* \end{bmatrix} \quad (3)$$

Where (.)* indicates the complex conjugate operation, and s is the space-time code with its columns representing space dimension while the rows representing the time dimension [14].
Alamouti encoded signal is transmitted from the two transmit antennas over two symbol periods. During the first symbol period, two symbols $s_1$ and $s_2$ are simultaneously transmitted from the two transmit antennas. During the second symbol period, these symbols are transmitted again, where $-s_2^*$ is transmitted from first transmit antenna and $s_1$* is transmitted from the second transmit antenna. With orthogonal STBC, $s_i$ and $s_i$* ($i = 1, 2$) have the same PAPR properties, and thus the PAPR reduction needs to be done only for the first symbol period [15].

## 3. The PAPR of the signal:

The PAPR of the signal, *x(t)*, is then given as the ratio of the peak instantaneous power to the average power, written as

$$PAPR = \frac{\max_{0 \leq t \leq T} |x(t)|^2}{E[|x(t)^2|]} \quad (4)$$

Where $E[\cdot]$ is the expectation operator. From the central limit theorem, for large values of N, the real and imaginary values of *x(t)* becomes Gaussian distributed. The amplitude of the OFDM signal, therefore, has a Rayleigh distribution with zero mean and a variance of N times the variance of one complex sinusoid.

In practice, they calculate the probability of PAPR exceeding a threshold as measurement index to represent the distribution of PAPR. This can be described as "Complementary Cumulative Distribution Function" (CCDF) and its mathematical expression as [16]

$$CCDF(PAPR(x(n))) = P_r(PAPR(x(n))) > PAPR_0 \quad (5)$$

Due to the independence of the N samples, the CCDF of the PAPR of a data block with Nyquist rate sampling is given by

$$P = P_r(PAPR(x(n))) > PAPR_0 = 1 - (e^{-PAPR_0})^N \quad (6)$$

This expression assumes that the N time domain signal samples are mutually independent and uncorrelated and it is not accurate for a small number of subcarriers. Therefore, there have been many attempts to derive more accurate distribution of PAPR [16].

There have been many attempts to find the exact probability distribution of PAPR for analysis MIMO-OFDM system PAPR performance is the same as if each SISO single antenna. For the entire system, the PAPR is defined as the maximum among all the PAPR transmit antennas from [17]:

$$PAPR_{MIMO-OFDM} = \max_{i \leq i \leq n_t} PAPR_i \quad (7)$$

Where PAPRi denotes the PAPR of transmit antenna. Specifically, since in MIMO-OFDM, Mt N time domain samples are considered compared to N in the SISO-OFDM system, the CCDF of PAPR in MIMO-OFDM is written as follows [17]:

$$P_r(PAPR_{MIMO-OFDM} > PAPR_0 = (1 - e^{-PAPR_0})^{M_T N} \quad (8)$$

## 4. Techniques for PAPR reduction:

### 4.1. Clipping and filtering:

One of the methods used for eliminate this high peaks is clipping and filtering method. The OFDM signal is deliberately clipped at a particular threshold value before amplification in this method [18]. The large peaks of OFDM signals are arose with a very low probability and hence clipping could be an effective technique for the reduction of the PAPR.

However, clipping cause important in-band distortion and out-of-band noise which will indirectly degrades the bit error rate performance and the spectral efficiency. Direct clipping suppresses the time-domain OFDM signals of which the signal powers exceed a certain threshold. The penalty is the significant increase of out-of-band energy. Peak windowing or filtering after direct clipping can be used to reduce the outof- band energy. After the filtering operation, the peak of the time-domain signal may increase. Hence, recursive clipping and filtering (RCF) can be used to suppress both the out-of-band energy and the PAPR. RCF can be modified by restricting the region of distortion to obtain improved error performance, filtering is done after clipping in order to eliminate unwanted frequencies caused by the clipping process [19].
Clipping and filtering algorithm for OFDM transmitter block diagram as shown in figure 2.

The main idea of Clipping and Filtering algorithm is to limit distortion of the frequency domain to approximate estimate and processing. The processing steps are as follows:

a) The frequency domain signal through the IFFT transform, received oversampling time-domain signal.
b) Clipping in the time domain, and then clipping distortion to the frequency domain by FFT transform.
c) Out-of-band signal is set to zero, artificially.
d) Using IFFT converted to time domain signal, and output.

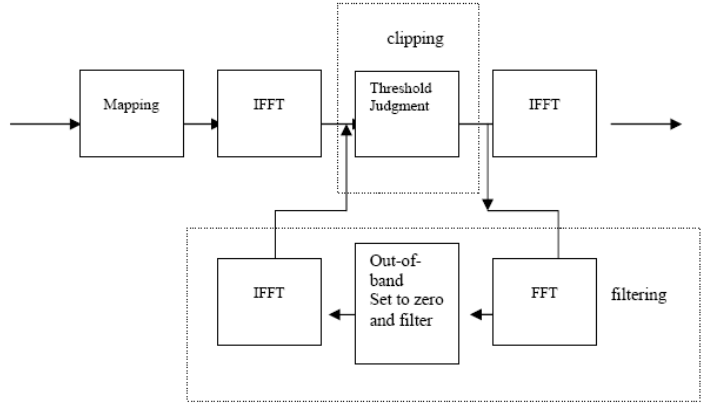

Figure.2 Clipping and filtering algorithm for OFDM transmitter block diagram

## 4.2. Selective mapping method (SLM):

Another solution is to use selective mapping method (SLM): The entire data stream is divided into different blocks of N symbols each. Each block is multiplied with U different phase factors to generate U modified blocks before giving to IFFT block. Each modified block is given to different IFFT block to generate OFDM symbols. PAPR is calculated for each modified block and select the block which is having minimum PAPR ratio. This technique can reduce PAPR considerably. But this technique will increase circuit complexity since it contains several IFFT calculations [1].

Let.s define data stream after serial to parallel conversion as X= $[X_n = X_0, X_1, \ldots \ldots X_{N-1}]^T$.

Initially each input $X_n^{(u)}$ can be defined as equation

$$x_n^{(n)} = x_n . b_n^{(n)} \qquad (9)$$

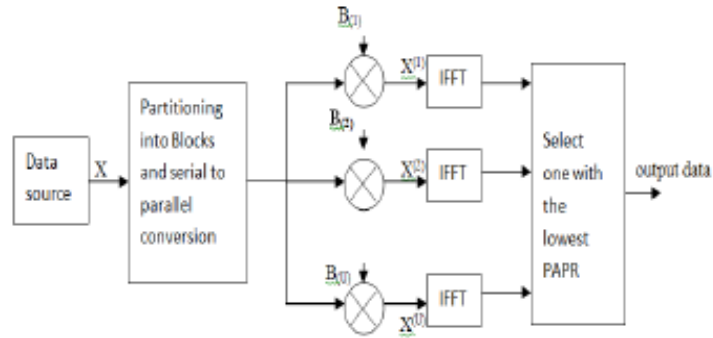

Figure 3. Block Diagram of OFDM transmitter with the SLM Technique [1]

$B^{(u)}$ can be written a $x_n^{(u)} = \left[x_0^{(u)} x_1^{(u)}, \ldots \ldots x_{N-2}^{(u)}\right]^T$. Where n= 0, 1, 2………N-1, And u=0,1,2...U to make the U phase rotated OFDM data blocks. All U phase rotated OFDM data blocks represented the same information as the unmodified OFDM data block provided that the phase sequence is known [20].

After applying the SLM technique, the complex envelope of the transmitted OFDM signal becomes:

$$x(t) = \frac{1}{\sqrt{N}} \sum_{n=0}^{N-1} x_n e^{j2\pi n \Delta f t}, 0 \le t \le NT \qquad (10)$$

Where $\Delta f = \frac{1}{NT}$, NT is the duration of an OFDM block.

Output data of the lowest PAPR is selected to transmit PAPR reduction effect will be better as the copy block number U is increased. SLM method effectively reduces PAPR without any signal distortion. But it has higher system complexity and computational burden. This complexity can less by reducing the number of IFFT block [21, 22, 23].

### 4.3. Partial transmit sequence technique (PTS):

In PTS technique, an input data block of N symbols is partitioned into disjoint subblocks. The subcarriers in each subblock are multiplied by a phase factor.

Figure 4 [24] is the block diagram of PTS technique. From the left side of diagram, the data information in frequency domain X is separated into V non-overlapping sub-blocks and each subblock vectors has the same size N. So for each and every sub-block it contains N/V nonzero elements and set the rest part to zero. Assume that these sub-blocks have the same size and no gap between each other [25]. The sub-block vector is given by

$$X = \sum_{v}^{V} b_v X_v \qquad (11)$$

Where $b_v = e^{j\phi_v}$ ($\phi_v \in [0, 2\pi]$) { $v = 1, 2, \ldots, X_v$ }

is a weighting factor been used for phase rotation.

The signal in time domain is obtained by applying IFFT operation on, that is

$$\hat{x} = IFFT(X) = \sum_{v=1}^{V} b_v IFFT(X_v) = \sum_{v=1}^{V} b_v X_v \quad (12)$$

For the optimum result one of the suitable factor from combination b = [$b_1$, $b_2$,.., $b_v$] is selected and the combination is given by

$$b = [b_1, b_2, ......, b_v] = \arg\min_{(b_1, b_2......b_v)} (\max_{1 \leq n \leq N} \left|\sum_{v=1}^{V} b_v X_v\right|^2) \quad (13)$$

Where arg min [(·)] is the condition that minimize the output value of function.

The phase factors are selected such that the PAPR of the subblocks is minimized. Each of the subblocks having the minimum PAPR and hence the combined signal of the different subblocks is having the minimized PAPR.

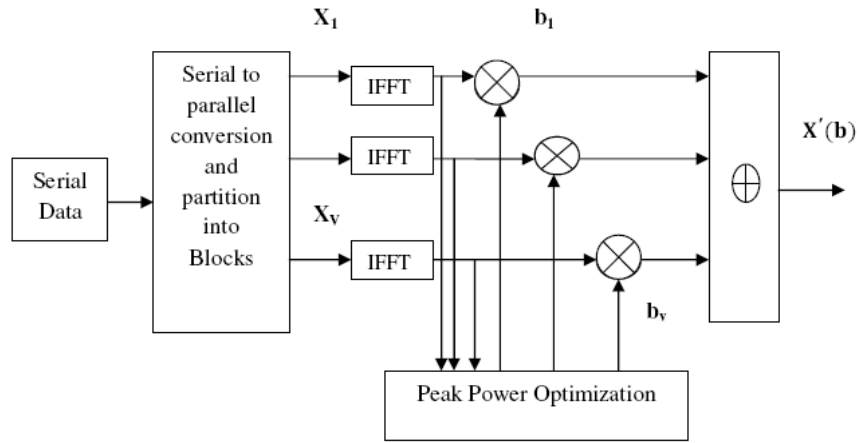

Figure 4. The Block diagram of PTS Technique [21]

## 5. Simulation analyses of the PAPR reduction:

To implement active set approach for PAPR reduction signal we generate OFDM symbols of length 512 and 1024 samples with 301 and 601 used subcarriers respectively. Techniques have been done using MATLAB 7.10. The simulation parameters considered for this analysis are summarized in Table 1.

| Parameters | Values |
| --- | --- |
| System Sub-carrier | 512,1024 |
| MIMO Scheme (Tx ×Rx) | (2×2) |
| Oversampling factor (L) | 6 |
| Modulation scheme | QPSK |
| Phase factor | 1,-1, j, -j |
| Clipping Ratio (CR) | 4 |

| Route numbers used in SLM method | M =8 |
| Number of sub-blocks used in PTS methods | V=8 |
| Random OFDM symbols generated | 1000 |

Table 1. Parameters used in clipping and filtering, PTS, SLM and algorithm

Table 1 show the parameters of OFDM signal which is used for PAPR reduction. Here, the number of sub-carriers used are N=512, 1024, An MIMO-OFDM system modulated with QPSK is used for the simulation. The complementary cumulative distribution function (CCDF) of the PAPR for the transmitted signal are plotted after each methods reducing PAPA used.
The flow charts used for PAPR reduction technique is given in Figure 5.

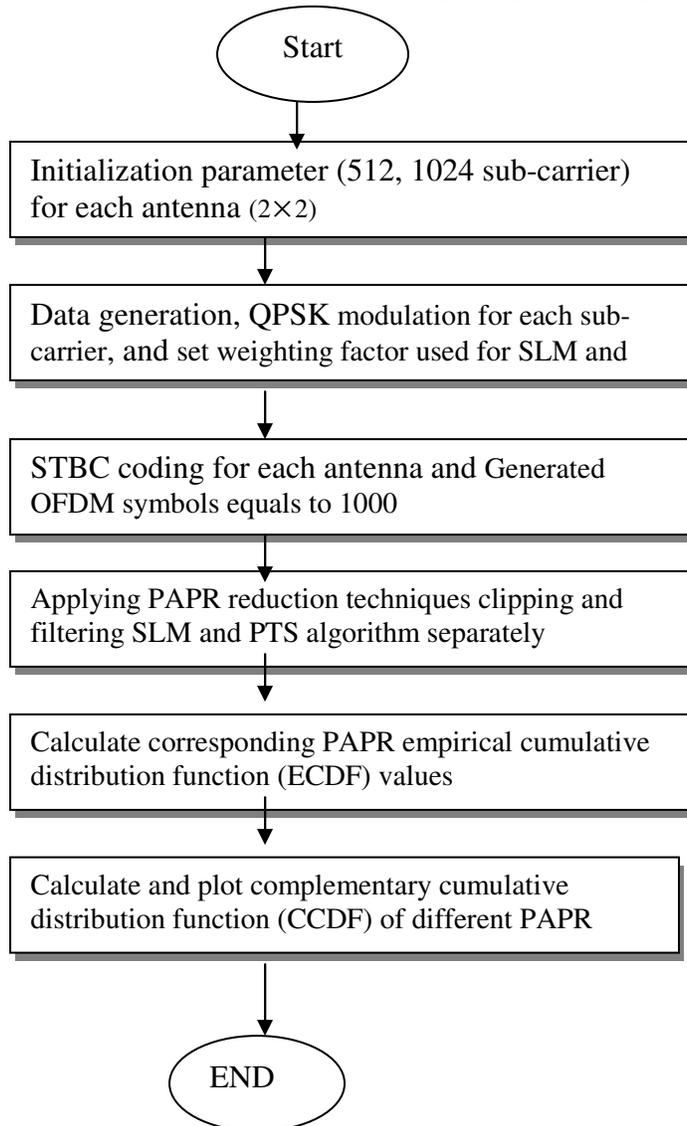

Figure 5. PAPR Reduction algorithm techniques clipping and filtering SLM and PTS

## 5. Simulation Results:

The results of the simulations are presented in this section. To implement the PAPR reduction of a signal MIMO-OFDM( STBC) we generated OFDM symbols of length 512and 1024 samples 301and 601 tones are used for data transmission and PAPR reduction. Each of the data carrying tones uses a QPSK modulation, an oversampling factor of L = 6 is applied and in our simulation 1000 randomly selected tone sets are generated.

The results of PAR reduction in the simulations are presented as the Complementary Cumulative Density Function (CCDF) of the PAPR of the STBC MIMO-OFDM signals.

The Figures 6,7,8 shows the magnitude of the peak reduction symbol of length 512 with different methods clipping and filtering, partial transmit and select mapping respectively for a system (2x2) MIMO-OFDM ( STBC) . We see that the reduction in clipping and filtering is nearly 4 dB, while in some literatures art clipping and filtering achieved only a PAPR reduction of 2dB. This confirms that the STBC coding helps in ways beneficial to reduce fluctuations in the envelope of the MIMO-OFDM with association generally block codes space-system; it takes advantage of the spatial diversity obtained by spatially separated antennas.

In the others methods PTS and SLM reduction of PAPR does not exceed 1.5dB and 2dB respectively

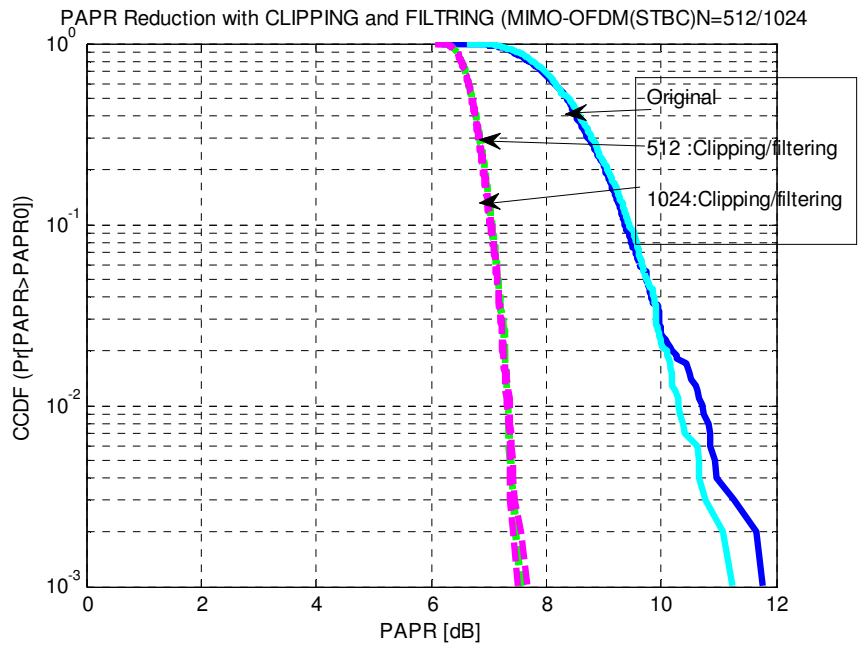

Figure6. PAPR Reduction with clipping and filtering STBC MIMO-OFDM (N=512/1024).

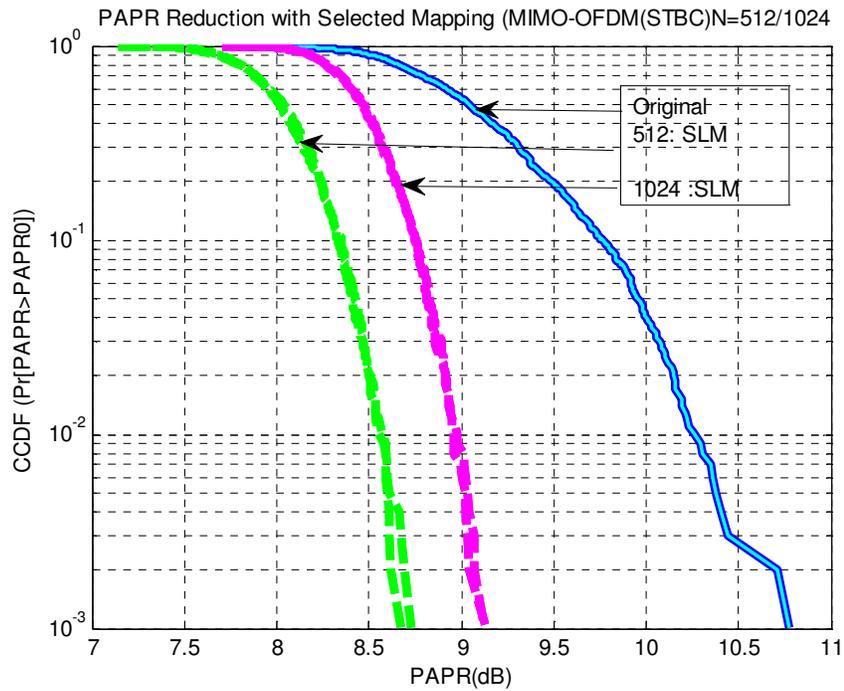

Figure7. PAPR Reduction with selected mapping STBC MIMO-OFDM (N=512/1024)), M=8

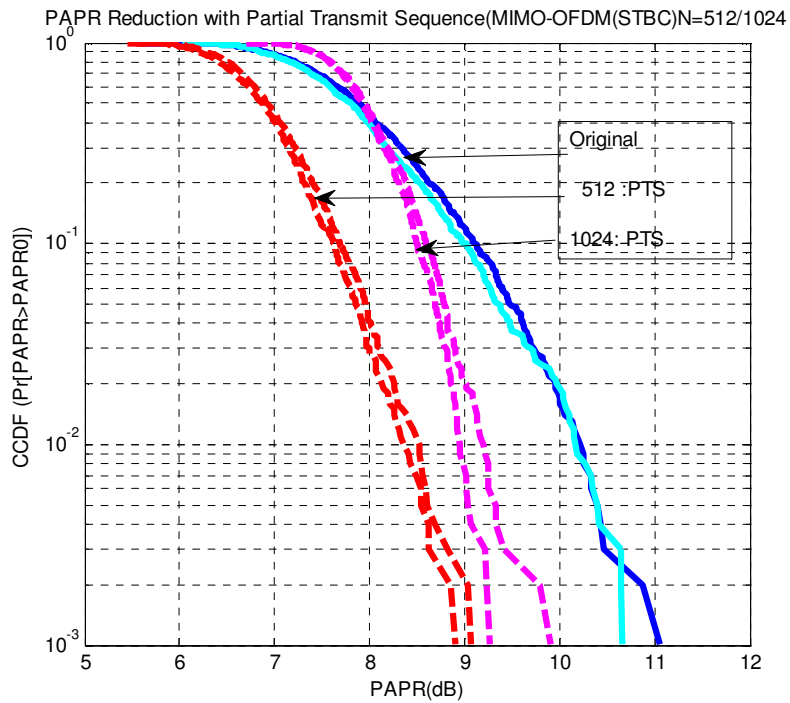

Figure8. PAPR Reduction with Partial Transmit Sequence STBC MIMO-OFDM (N=512/1024), V=8

In the second part a performance comparison between different methods of the PAPR reductions is done to distinguish which technique conserve the reduction found in the transmission

Figure 9,10 shows the performance in terms of CCDF of PAPR of the signal received, a comparison between different methods of the PAPR reductions is done.While the clipping and filtering method shows a slight decrease in PAPR, PTS gives results which diverges from original curves bat with SLM technique PAPR reduction result exceed 2 dB it is equivalent to reduction PAPR calculate in the transmission part. SLM not only reduces the complexity at the reception, but it also reduces the PAPR of the OFDM signal [22]

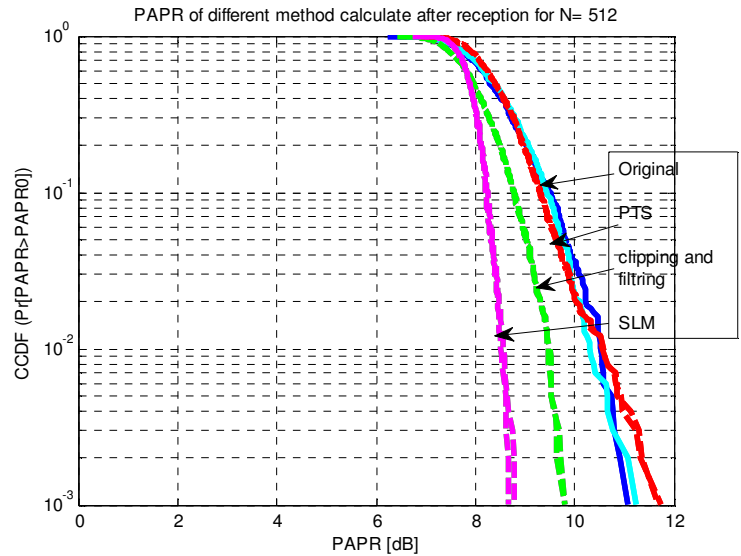

Figure9. PAPR of different methods STBC MIMO-OFDM (N=512), V=8, M=8

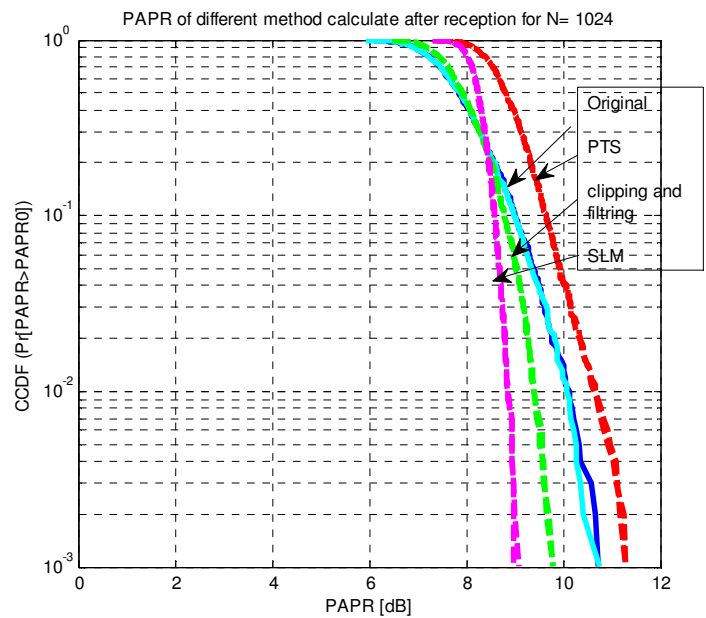

Figure10. PAPR of different methods STBC MIMO-OFDM (N=1024), V=8, M=8

## 6. CONCLUSION:

In this paper, we present in the first part an analysis of the PAPR reduction method which are clipping and filtering, PTS, SLM in STBC MIMO-OFDM system. Simulation results have shown that clipping and filtering technique gives a better reduction PAPR (4 dB) compared to the others methods. In the second part a result represent by curves CCDF of PAPR of signal received into different methods exhibit a conservation of the PAPR reduction in SLM technique.

Systems" 16th international workshop on computer aided modeling and design of communication links and networks (CAMAD) IEEE 2011

[16] P. Mukunthan and P.Dananjayan, "Modified PTS with FECs for PAPR Reduction of OFDM Signals," International Journal of Computer Applications, Vol. 11, No.3, pp. 38-43, December 2010.

[17] P.Mukunthan, P.dananjayan, "Modified PTS with FECs for PAPR redaction in MIMO-OFDM system with different subcarriers," international symposium of humanities, Science and engineering Research IEEE 2011

[18] Seung Hee Han and Jae Hong Lee, "Modified selected Mapping Technique for PAPR reduction of coded OFDM signal," IEEE Transaction on broadcasting, Vol. 50, No.3, pp.335-341, Sept 2004.

[19] Pawan Sharma, Seema Verma "PAPR reduction of OFDM signals using selective mapping with turbo codes" International Journal of Wireless & Mobile Networks (IJWMN) Vol. 3, No. 4, August 2011

[20] Yang Jie, Chen Lei, Liu Quan and Chan De, "A Modified selected mapping technique to reduce the Peak to Average Power Ratio of OFDM signal," IEEE transaction on consumer Electronics, Vol53, No.3, pp. 846-851, August 2007.

[21] Stefan H.Muller and Johannes B. Huber,"A Comparison of Peak Power Reduction Schemes for OFDM," In Proc. of The IEEE Global Telecommunications conference GLOBECOM. 97, Phonix, Arizona, USA, pp.1-5, Nov. 1997.

[22] Marco Breiling ,Stefan H. Muller-Weinfurtner and Johanes B.Huber, "SLM Peak-Power Reduction Without Explicit Side Information", IEEE Communications Letters,Vol. 5, No.6, pp.239-241, JUNE 2001.

[23] Jayalath, A.D.S, Tellainbura, C, "Side Information in PAR Reduced PTS-OFDM Signals," Proceedings 14th IEEE Conference on Personal, Indoor and Mobile Radio Communications, Vol.1, Sept 2003.

[24] Oh-Ju Kwon and Yeong-Ho Ha, "Multi-carrier PAP reduction method using sub-optimal PTS with threshold," IEEE Transactions on Broadcasting, vol. 49, June 2003.

[25] Suverna Sengar1, Partha Pratim Bhattacharya, "Performance Improvement in OFDM system by papr reduction", Signal & Image Processing: An International Journal (SIPIJ) Vol.3, No.2, April 2012.